\documentclass[preprint,showpacs,preprintnumbers,superscriptaddress,amsmath,amssymb]{revtex4}

\usepackage{graphicx}% Include figure files
\usepackage{dcolumn}% Align table columns on decimal point
\usepackage{bm}% bold math

\begin{document} 

\title{Spallation nuclei in {substellar objects}: a new dark-matter signature?} 
\author{Benjamin Monreal}
\affiliation{Laboratory for Nuclear Science and Department of Physics, Massachusetts Institute of Technology, Cambridge MA 02139}
\email{bmonreal@mit.edu}
\author{Lorne A. Nelson}
\affiliation{Department of Physics, Bishop's University, Sherbrooke QC}
\author{Joseph A. Formaggio}
\affiliation{Laboratory for Nuclear Science and Department of Physics, Massachusetts Institute of Technology, Cambridge MA 02139}
\date{\today}

\date{\today}

\begin{abstract}Although dark matter makes up 80\% of the gravitational mass of our Galaxy, its composition is not known. One hypothesis is that dark 
matter consists of massive particles called WIMPs.  WIMPs are 
expected to accumulate and coannihilate in the cores of stars, but 
the only signature of this accumulation has been thought to be hard-
to-observe high-energy neutrinos. Here we propose an entirely new 
observable signature.   WIMP coannihilations in the core of a very 
low-mass star, brown dwarf, or planetary-mass object should alter the 
star's chemical composition via spallation reactions.  Very close to 
the Galactic center, these stars may acquire extremely high lithium, 
beryllium, and boron abundances, even for models with otherwise-
undetectable WIMP-nucleon cross sections.   These abundances should 
be measurable in certain stellar systems and phenomena.\end{abstract}


\maketitle

\section{Introduction}
In this era of high precision cosmology, the mass-energy density of the Universe is well-constrained.  {From a combination of measurements of the microwave background, large-scale strucure, rotation curves, gravitational lensing, and galaxy cluster temperatures, the Universe is known to contain about 4\% ordinary matter and 24\% dark matter.}  Dark matter is known to move and gravitate just like massive baryonic matter, but without any particle-particle interactions at the electromagnetic or strong-force scales.  The Standard Model of particle physics contains no massive non-interacting particle like this, but some extended models do; such particles are called WIMPs.  Supersymmetry (SUSY) naturally includes a WIMP, the neutralino ($\chi_0$), which is an excellent dark-matter candidate; Kaluza-Klein bosons, which arise in theories with large or warped extra dimensions, are another example.  In this work we refer to WIMPs as $\chi_0$, but our results are not model-specific (except as discussed below).    

In order for WIMPs to be produced in the Big Bang, there must be a sizeable cross section for reactions like $x\bar{x} \rightarrow \chi_0\chi_0$, where $x$ is some Standard Model particle.   This implies that the reverse reaction, $\chi_0\chi_0 \rightarrow x\bar{x}$, will occur today at a rate proportional to the square of the WIMP density.  The WIMP-nucleus elastic scattering cross section, $\chi_0 ^aN_z \rightarrow \chi_0 ^aN_z$ 
is not constrained by astrophysics, and may be extremely small.  
Terrestrial experiments allow us to constrain this cross section to be $\sigma_0/A^2 \tilde{<} 10^{-43}$ cm$^2$ for spin-independent WIMP-nucleus scattering, and $\sigma_p \tilde{<} 10^{-36}$ cm$^2$ for spin-dependent WIMP-proton scattering.

Any nonzero cross section for $\chi_0 ^aN_z \rightarrow \chi_0 ^aN_z$ 
% LAN {allow a Galactic $\chi_0$s to scatter off nuclei in stars, to enter bound orbits which cross the star repeatedly, and thus to thermalize}
{allows} a Galactic {$\chi_0$ particle} to scatter off nuclei in stars, to enter bound orbits which cross the star repeatedly, and thus to thermalize
and be captured
inside the star.  As the $\chi_0$ number density in the star's core increases, the coannihilation rate will increase until it equilibrates with the  capture rate.  Recent work\cite{moskalenko_dark_2007} shows that, in regions near the Galactic 
center's SuperMassive Black Hole (SMBH)
 where the dark-matter density is highest, the capture-and-annihilation process may provide enough thermal power to change a star's luminosity and evolution.  We show that these coannihilations can change stellar \emph{chemical} composition in unambiguous ways.  Most importantly, the high-energy annihilation products can convert carbon and
%BM Lets call them SSOs.   Define them here, use "SSO" throughout except for conclusion section.
% LAN -- agreed.  Note the change to" $M < 0.1 M$_\odot$ on the next line and the two sets of terms in {}
oxygen into $^3$He, Li, Be, and B via spallation.  We restrict our analysis to stars with $M < 0.1$ M$_\odot$ (encompassing {the lowest-mass} main-sequence stars, brown dwarfs, and planetary-mass objects) which {we will refer to loosely as} substellar objects (SSOs).

\section{Spallation reactions and WIMP capture}

\begin{figure}
\begin{center}
\includegraphics[width=5in]{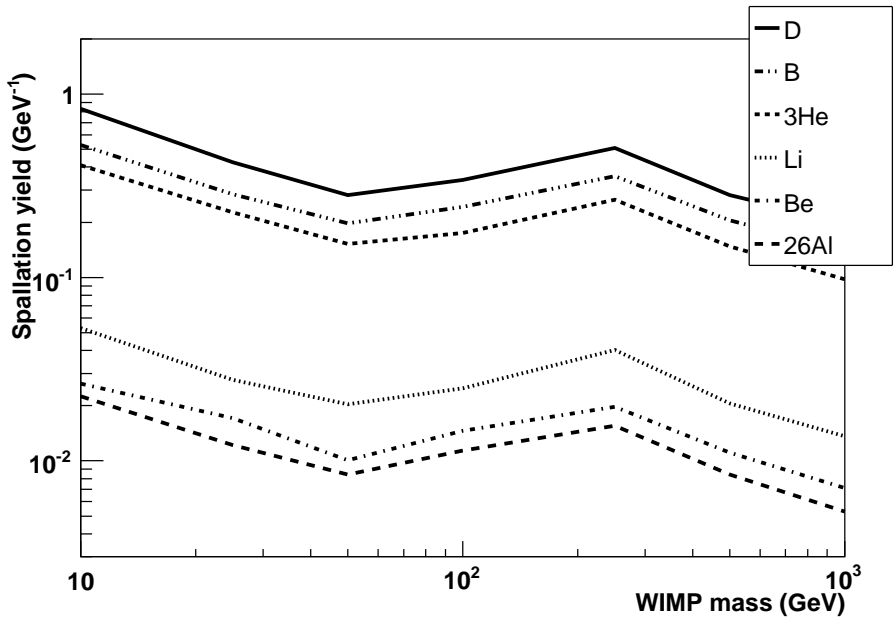} 
\end{center}
\caption{Spallation nuclei produced per primary GeV, based on simulations of b-, W-, or t-jets incident on degenerate SSO core matter.   
% LAN  see the term in {} on the next line
Yields for mixed solar material are consistently a factor of {$\sim 100$ lower.}} \label{figSpall}
\end{figure}

%BM clarify with intro sentence
% LAN  -- good  Small change in {} on next line
In order to quantify the spallation yields, we {perform} a numerical simulation of WIMP coannihilation products colliding with stellar core matter.  Many SUSY WIMP models suggest that $\chi_0\chi_0 \rightarrow x\bar{x}$ usually proceeds, via prompt heavy quark or W boson pairs, into hadron jets containing 50--100 particles.  Using the Pythia6 event generator\cite{sjostrand_pythia_2006}, we create WIMPs at rest and annihilate them to the heaviest allowed of $b\bar{b}$, $W^+W^-$, or $t\bar{t}$\footnote{This is a reasonable assumption but not a universal one\cite{jungman-1996-267}.   SUSY model WIMPs do not generally annihilate to light quarks or to leptons, for which the spallation yields would be much lower; $\tau\bar{\tau}$ occurs in some parts of parameter space, but differs in yield from $b\bar{b}$ only by a factor of a few.  Kaluza-Klein dark matter annihilates to heavy as well as light final states\cite{bringmann_high-energy_2005}, suppressing spallation yields by a factor of a few.}
% LAN  -- several changes follow.  All are enclosed in {}
% We use the Geant4 Monte Carlo package\cite{agostinelli_geant4-a_2003} to propagate the final states through 
%BM - add note on mixed material
%two relevant materials: uniform solar material (Z=1--26), and the approximate core composition of a gravitationally-settled star (solar composition with Z=6--26)\footnote{To account for the dense, degenerate matter, we inhibit ionization losses; we approximate $\pi^0$ hadronic interactions as being identical to $\pi^-$, but otherwise use the QGSP Bertini nuclear model.}.  The spallation yield per unit $\chi_0$ mass is shown in Figure \ref{figSpall} for the core material; yields for uniform material are a factor of 100 lower.  
{.}  We use the Geant4 Monte Carlo package\cite{agostinelli_geant4-a_2003} to propagate the final states through 
two relevant {compositions: (i) a} uniform solar {composition ($Z=$1--26); and, (ii)} the approximate core composition of a gravitationally-{differentiated} star (solar composition with {$Z=$}6--26)\footnote{To account for the dense, {(partially)} degenerate matter, we inhibit ionization losses; we approximate $\pi^0$ hadronic interactions as being identical to $\pi^-$, but otherwise use the QGSP Bertini nuclear model.}.  The spallation yield per unit $\chi_0$ mass is shown in Figure \ref{figSpall} for the core material{. Note that the yields for a fully mixed composition} are a factor of 100 lower.

Next, we derive the WIMP capture rate for a range of example stars near the Galactic center, using 
{the analytical formula (equation 2.27) derived in} \cite{gould_resonant_1987}.   We assume that the Galactic halo has an adiabatic central ``spike'' with a static density profile of $\rho(r) = 10^5 $M$_\odot$/pc$ \times r^{-2}$.   Many dark-matter halo profiles are assumed to have a spike\cite{gondolo_dark_1999} as long as it has not been disrupted by a
%LAN {} on next line
galaxy merger; our assumption is consistent with, but not required by, {the} observations.   We have not accounted for the evolution of the spike, whose density would have been higher earlier in its formation history\cite{bertone_time-dependent_2005}.  (To adapt our results to other halo/spike models, the results may be scaled linearly with WIMP density.)  
% LAN  The previous sentence was put into square brackets (BM-disagree) and there are three changes to the next sentence in {}.
{To simplify the calculations, we} treat the stars as 
%BM mention circular orbits; this is not a trivial assumption!    
$n=1.5$ polytropes in circular Galactic orbits.  There are two distinct possibilities for the WIMP elastic scattering cross section: {(i)} spin-dependent{; and, (ii)} spin-independent.   Spin-independent capture on 
SSOs is very inefficient, since they consist mostly of light elements while the cross section scales as 
the nuclear mass squared.  However, the spin-dependent interaction may permit efficient scattering off {of} spin-1/2 
protons (hydrogen), allowing SSOs to be significant dark-matter burners.  The capture rate depends on the 
% LAN  Note the change in {} on the next line
SSO's mass ($M*$), WIMP mass ($M_{\chi_0}$), the {distance from the Galactic center} ($R$) and the WIMP-proton cross section.   
%BM This is a less-than, not a greater-than!   
As long as $\sigma_p < 10^{-36}$ cm$^2$, this rate simplifies to

% LAN The constants should probably be in Roman font (i.e., s, cm, pc, Msun, Rsun, GeV)   BM - OK, done
%\input{fit_jupsd2.tex}
\begin{equation}
C = \frac{10^{ 33.6 }}{\mathrm{s}} (\frac{\sigma_p}{10^{-36} \mathrm{ cm}^2})^{ 1 } (\frac{r}{\mathrm{pc}})^{ -1/2 }  (\frac{M*}{M_\odot})^{ 3 }   (\frac{R*}{R_\odot})^{ -2 } (\frac{M_{\chi_0}}{100 \mathrm{ GeV}})^{ -2 } 
\end{equation}

\noindent The capture rate is independent of $\sigma_p$ for $\sigma_p > 10^{-36}$ cm$^2$.

\section{Spallation of substellar object cores} 

If the WIMPs are in thermal equilibrium with the star, they will be collect near the star's core.   The creation of spallation products
and the production of energy will also be largely confined to the core region.  This has important implications both for the thermal\cite{moskalenko_dark_2007} and the chemical structure of these stars.   

In a uniformly-mixed SSO, spallation would occur mainly on H and He, with B production on C and O being suppressed.   %However, many SSOs have had their metal content settle gravitationally into a central core\cite{dantona_evolution_1985}; this core is large enough to enclose the entire WIMP annihilation region.   In these settled stars, the spallation target will be mostly C and O (producing abundant Li, Be, and B), with traces of elements up to Fe (producing rare elements like Sc, V, and radioactive $^{26}$Al and $^{40}$K).  Nuclear re-burning of the spallation products is negligible for stars with $M < 0.1 $M$_\odot$.  Stars with 0.1 < M < 0.3 can burn Li and Be but not B, and stars with M > 0.3 can burn B.   
% LAN several suggested changes in {}
However, many SSOs have {likely experienced chemical differentiation wherein most of their metals would have gravitationally settled} into a central core\cite{dantona_evolution_1985}; this core is large enough to enclose the entire WIMP annihilation region.   In these {differentiated} stars, the spallation target will be mostly C and O (producing abundant Li, Be, and B), with traces of elements up to Fe (producing rare elements like Sc, V, and radioactive $^{26}$Al and $^{40}$K).  Nuclear re-burning of the spallation products is negligible for stars with $M < 0.1 $M$_\odot$.  Stars with {$0.07 < M/$M$_\odot < 0.1$} can burn Li and Be but not B, and stars with $M > 0.1$M$_\odot$ can burn B\cite{nelson_1993}.
%LAN Note the new citation: {nelson_1993} L.A. Nelson, S. Rappaport, & E. Chiang, ApJ, 413, 364-367 (1993).  
We note that, unlike a thermal signature, a chemical signature may persist indefinitely for a star which has been \emph{ejected} from the Galactic center and has ceased burning WIMPs.      
  
For concreteness, we define a set of 
representative substellar objects and calculate their expected WIMP annihilation rates and spallation accumulation rates for a WIMP mass of $M_{\chi_0} = 100$ GeV and a spin-dependent WIMP-proton cross section of $\sigma_0 = 10^{-38}$ cm$^2$.
%LAN Note the {} insertion.
The results are {shown} in Table \ref{tabModels}.  

\begin{table}
 \begin{tabular}{|c|c|c|c|c|c|c|}
 \hline
Exam-  & Mass & radius     & $\chi_0$ rate & power &  \multicolumn{2}{|c|}{Boron atom abund./10 Gyr}\\ 
ple \#     &(M$_\odot$) & (pc)  &     s$^{-1}$         &  (L$_\odot$)  & metal core & mixed metals\\
 \hline 
 \hline 
1 &  $10^{-1}$ &  $10^{-2}$ & $10^{32}$ & $10^{-2.8}$  & -- & $10^{-8.3}$ \\ 
2 &  $10^{-1}$ &  $10^{-5.5}$ & $10^{33}$ & $10^{-1.1}$  & -- & $10^{-6.5}$ \\ 
3 &  $10^{-1.2}$ &  $10^{-1}$ & $10^{31}$ & $10^{-3.5}$  & $10^{-6.8}$ &  $10^{-8.8}$ \\ 
4 &  $10^{-1.2}$ &  $10^{-2}$ & $10^{31}$ & $10^{-3}$  & $10^{-6.3}$ &  $10^{-8.3}$ \\ 
5 &  $10^{-1.2}$ &  $10^{-5.5}$ & $10^{33}$ & $10^{-1.3}$  & $10^{-4.5}$ &  $10^{-6.5}$ \\ 
6 &  $10^{-2}$ &  $10^{-2}$ & $10^{29}$ & $10^{-5.8}$  & $10^{-8.3}$ &  $10^{-10}$ \\ 
7 &  $10^{-2}$ &  $10^{-5.5}$ & $10^{30}$ & $10^{-4.1}$  & $10^{-6.5}$ &  $10^{-8.5}$ \\ 
8 &  $10^{-3}$ &  $10^{-2}$ & $10^{26}$ & $10^{-8.8}$  & $10^{-10}$ &  $10^{-12}$ \\ 
9 &  $10^{-3}$ &  $10^{-5.5}$ & $10^{27}$ & $10^{-7.1}$  & $10^{-8.5}$ &  $10^{-11}$ \\ 
 \hline 
\end{tabular} 

\caption{
%LAN  Note the {}'s in this caption.
Seven examples of Galactic-center {SSOs}.  We compute the spallation-product contents for an arbitrary dark matter model with $m_{\chi_0}$ = 100 GeV and spin-dependent $\sigma_p$ = 10$^{-38}$.  We calculate the thermal power and capture/burning rate at equilibrium.  We {also show the average boron abundance of a fully mixed SSO} after 10 Gyr, first under the {\emph assumption} that the {SSO is gravitationally settled} and second under the {\emph assumption} that it is {fully} mixed. 
}\label{tabModels}
\end{table}      

\section{Observational prospects and constraints }
\begin{figure}
\begin{center}
\includegraphics[width=5in]{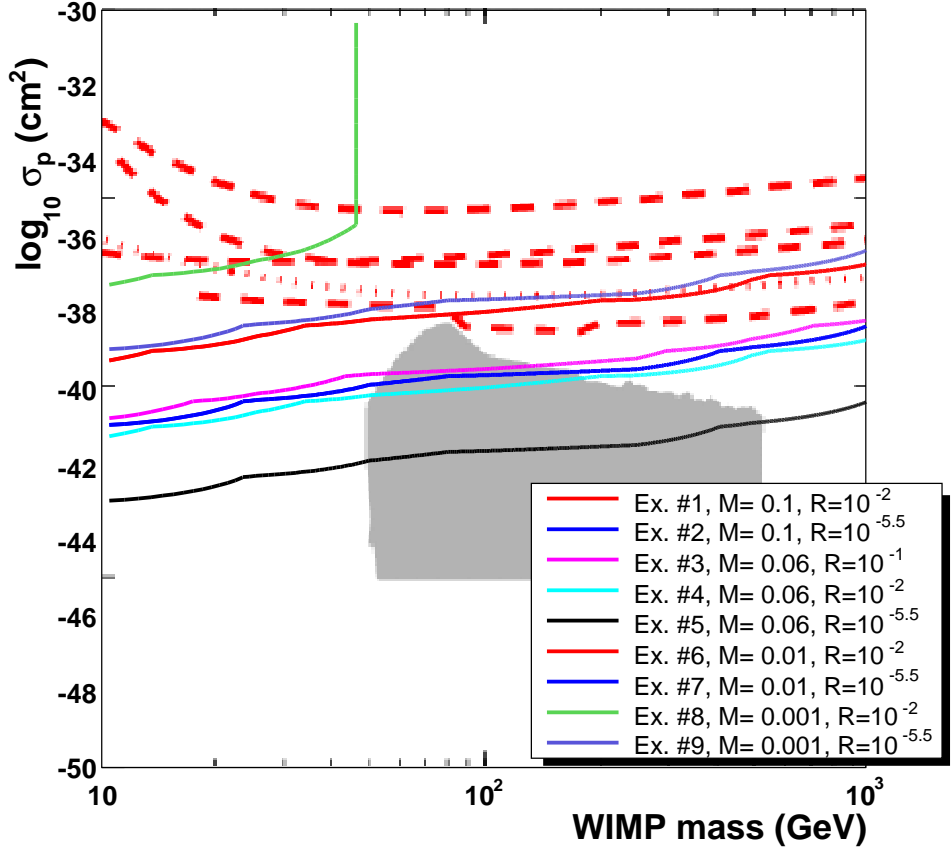}
\end{center}
\caption
% LAN  The caption changes in {}
{Discovery potential for dark matter in observations of spallation products.  The graph shows the spin-dependent $\chi_0$-proton elastic scattering cross section versus the WIMP mass.  The colored solid contours show where the example stars   {defined in Table \ref{tabModels} would achieve a total B/H ratio $ > 10^{-8}$ (quantities in {the} legend are {the} mass in M$_\odot$ and Galactic radius in {pc}).} 
%BM
Examples 1 and 2 {assume complete mixing}, all others assume full gravitational settling.   Dashed (dotted) red lines are existing (proposed) experimental limits; top to bottom, {they are:} CDMS\cite{akerib_limits_2006}, KIMS\cite{lee_limits_2007}, COUPP\cite{coupp_2007}, NAIAD\cite{spooner_nai_2000} (100 kg-yr), and Super-Kamiokande\cite{desai_2004}.   {A family of MSSM models is denoted by the shaded-grey region}.   Limits and SUSY {data are taken} from \cite{gaitskell_dark_}.}
\label{figLimits2}
\end{figure}

%LAN A couple more {} in the next 2 sentences.
Low-background nuclear physics experiments like CDMS
\cite{akerib_limits_2006} typically plot their dark-matter discovery potential as a function of $ \sigma_p$ and {$M_{\chi _0}$}. We do the same for spallation production in SSOs in Figure \ref{figLimits2}. The colored contours show where the example stars from Table \ref{tabModels} can accumulate a net boron {atomic} abundance of $ > 10 ^{-8}$ in {10 Gyr}. 
Our results demonstrate that extremely low-mass stars, brown dwarfs, and planetary mass objects can produce anomalous abundances of Li, Be, and B if: (1) these objects reside within the inner parsec of the Galactic center; (2) the dark matter consists of WIMPs with properties as described; and, (3) the WIMP density distribution has a spike at the Galactic center. Very few other assumptions are required, except that WIMP coannihilation proceeds to hadrons. The discovery or exclusion of anomalous B levels in stars like these could provide a discovery channel for, or powerful constraints on, various dark matter models that might otherwise be inaccessible.
% LAN Suggested new sentence addition.
Moreover, any `positive identifications' that may determined from other experiments will need to be checked by as many \emph{independent} tests as possible. 

Unfortunately, the measurement of a detailed spectrum of even the brightest of these objects will be extremely challenging for any existing or foreseen telescope, due to the large distance (8 kpc), interstellar extinction, and crowding. However, some special cicumstances may allow us to observe the spallation byproducts of such objects in the future. For example, three-body encounters may be reasonably effiicient at ejecting low-mass objects from the deep gravitational potential well near the Galactic center and thus, on a gigayear timescale, seed the local environment with observable candidates. The detection of anomalous abundances of light elements such as boron in the atmospheres of isolated, very low-mass objects whose interiors are fully convective would be an important indicator of WIMP-induced spallation.

Interacting binaries containing compact accretors, such as Cataclysmic Variables and Low-Mass X-Ray Binaries, can also expose the chemical constituents of the interiors of their low-mass companions (i.e., the donor stars). Thus our candidate objects may serve as the donors in these systems (e.g., CV SDSS 103533.03+055158.4
\cite{littlefair_brown_2006}), and the accretion disks may make their composition visible in X-rays or the UV. 
% LAN  A couple more changes in {}
Moreover, some of these systems can cannibalize their companions to such a degree that only the cores of the donors persist (e.g., the black widow pulsar or {other} ultracompact binaries). These types of binaries allow us to potentially observe the spallation products of very low-mass objects whose interiors had been chemically differentiated and whose core material could not have been otherwise {dredged} up to the surface. 
% BM note superconcentration at core.
Moreover, if spallation products from the core have not mixed with the whole star, the core boron concentration may be orders of magnitude higher than the average concentrations discussed above.   
% LAN  A couple more changes in {}
Finally we note that the total disruption of a very low-mass object would give us the {possibility of observing} the core's chemical composition. In fact, one of our candidate objects may be disrupted by, and accrete onto, the SMBH itself (perhaps as in RX J1242-11 \cite{komossa_huge_2004}). The resulting burst may show evidence of spallation products. 
% LAN I changed the next sentence completely, but the original may be better.
% BM-how about this? 
Observation of the boron abundances of Galactic Center SSOs or SSO-related phenomenon, although difficult, should be a high priority for future UV missions.

The authors wish to thank Saul Rappaport, Adam Burgasser, and Gray Rybka.

\bibliographystyle{apsrev}
\bibliography{burners_prl}
   
\end{document}